\documentclass[aps,epsf,subfigure,twocolumn,floatfix]{revtex4}

\usepackage{graphicx}% Include figure files
\usepackage{dcolumn}% Align table columns on decimal point
\usepackage{bm}% bold math
\usepackage{amsmath}
\usepackage{subfigure}
\usepackage{color,psfrag,epsfig,colordvi}

\usepackage{color}
\usepackage{tabularx}
\usepackage{amssymb}
\usepackage{wasysym}
\usepackage{bm}
\usepackage{epstopdf}

\begin{document}

\newcommand{\up}{{\mid \uparrow \rangle}}
\newcommand{\down}{{\mid \downarrow \rangle}}

\title{Ar:N$_2$ - a non-universal glass}

\author{Alejandro Gaita-Ari\~no$^1$}
\author{Vicente F. Gonz\'alez-Albuixech$^1$}
\author{Moshe Schechter$^2$}
\affiliation{$^1$Instituto de Ciencia Molecular, Universidad de Valencia,
Cat. Jos\'e Beltr\'an, 2 46980, Paterna, Spain}
\affiliation{$^2$Department of Physics, Ben Gurion University of the Negev, Beer Sheva 84105, Israel}

\date{\today}

\begin{abstract}

The bias energies of various two-level systems (TLSs) and their strengths of
interactions with the strain are calculated for Ar:N$_2$ glass. Unlike the case
in KBr:CN, a distinct class of TLSs having weak interaction with the strain and
untypically small bias energies is not found.
The addition of CO molecules introduces CO flips which form such a class of
weakly interacting TLSs, albeit at much lower coupling than are typically
observed in solids. We conclude that because of the absence of a distinct class
of weakly interacting TLSs, Ar:N$_2$ is a non-universal glass, the first such
system in three dimensions and in ambient pressure. Our results further suggest
that Ar:N$_2$:CO may show universal properties, but at temperatures lower than
$\approx 0.1$ K, much smaller than typical temperature $\approx 3$ K associated
with universality, because of the untypical softness of this system. Our
results thus shed light on two long standing questions regarding low
temperature properties of glasses: the necessary and sufficient conditions for
quantitative universality of phonon attenuation, and what dictates the energy
scale of $\approx 3$ K below which universality it typically observed.

\end{abstract}

\maketitle

\section{Introduction}

One of the remarkable phenomena in condensed matter
physics is the universality of properties related to phonon attenuation at
temperatures smaller than $\approx 3$ K in a large class of materials, ranging
from amorphous solids, to disordered lattices, disordered polymers, and
quasi-crystals\cite{ZP71,HR86,PLT02}. The fact that universality is also
quantitative, both in the magnitude of phonon attenuation, and in the energy
scale dictating the temperature below which the phenomenon is observed, and the
broadness of systems exhibiting it, attests to the presence of a mechanism,
pertaining to the disordered state itself, that dictates phonon attenuation in
disordered systems. Exceptions to universality are rare, and have been observed
only in two-dimensional films under special conditions; hydrogenated silicon
films\cite{LWP+97} and silicon nitride films under applied stress\cite{SBV+09}.

Theoretically, much of the characteristics of disordered solids can be
understood within the ``standard tunneling model"
(STM)\cite{AHV72,Phi72,Jac72}, which introduces tunneling two-level systems
(TLSs); their interaction with the phonon field dominating phonon attenuation.
Within the STM all phonon attenuation properties are given in terms of the
``tunneling strength" $C_0 \approx 0.1 n \gamma^2/(\rho v^2)$, where n is the
density of states (DOS) of the TLSs, $\gamma$ is their interaction constant
with the phonon field (strain), $\rho$ is the mass density, and $v$ the
acoustic velocity. However, the STM can not explain what is the nature of the
TLSs, why is phonon attenuation universal among different systems, and the
origin of the energy scale of $\approx 3$ K dictating the universality regime.

Attempts to understand the nature of the TLSs include ``top-down" approaches,
trying to construct a theory which relies only on the glassy state of
matter\cite{YL88,Par94,BNOK98,LW01,Kuh03,PSG07}, and ``bottom-up" approaches,
attempting to identify the relevant TLSs in a given system, and then generalize
to all disordered systems showing universality\cite{SC85,YKMP86,GRS88,SK94}.
Specifically, the KBr:CN system was scrutinized, and it was suggested early
on\cite{SC85,SK94} that it is the flipping of the CN$^-$ impurities (and not
e.g. their rotations) that make for the relevant TLSs dictating the universal
properties in this system. To test this idea an experiment was carried on the
Ar:N$_2$ glass, which does not possess single impurities that can flip, with
and without the addition of CO impurities\cite{NYHC87}. The fact that the
linear term in the specific heat, corresponding to the DOS of the TLSs, had
little if any dependence on the CO concentration, was interpreted as a
refutation of the molecular flips as being the relevant TLSs dictating
universality. Little attention was paid to the fact that Ar:N$_2$ did not
exhibit universal phonon attenuation as observed in its thermal
conductivity\cite{YNH89}.

Recently, the fact that CN$^-$ flips in KBr:CN indeed constitute the TLSs
dictating universal phonon attenuation and the linear specific heat in KBr:CN
at $T \lesssim 3$ K received overwhelming support. Using Density Functional
Theory and ab-initio calculations it was shown that the coupling of CN$^-$
flips to the strain $\gamma_{\rm f}^{{\rm CN}^-} \approx 0.1$ eV, whereas the
coupling of CN$^-$ rotations to the strain $\gamma_{\rm r}^{{\rm CN}^-} \approx
3$ eV\cite{GS11}, the former value agreeing with the experimental value
obtained for the relevant TLSs at low temperatures\cite{BDL+85,YKMP86}.
Furthermore, the DOS of CN$^-$ flips and CN$^-$ rotations was calculated
numerically from first principles\cite{CBS14}, where it was shown that CN$^-$
flips are abundant at energies $\lesssim 3$ K, whereas CN$^-$ rotations are
scarce below $\approx 10$ K, as they are gapped by the weakly interacting
flips. All these findings are in excellent agreement with the recently
introduced two-TLS model\cite{SS09,CGBS13}, which derives the universality of
phonon attenuation and the energy scale of $\approx 3$ K for the universal
regime as a consequence of the existence of two classes of TLSs
differentiated by their interaction with the strain, based on the symmetry 
of the TLSs under inversion. Yet, the above mentioned results in KBr:CN and 
the two-TLS model seems to be at odds with the experimental results in
Ar:N$_2$:CO, as the latter suggest that the CO flips have no significant
contribution to the low energy properties of Ar:N$_2$:CO.

In this paper we reconcile this alleged discrepancy.
Using a discrete atomic model employing the Lennard Jones (LJ) potential, as
well as density function theory (DFT), we calculate the TLS-strain interaction
constant for various TLSs in both Ar:N$_2$ and Ar:N$_2$:CO systems. For the
latter we find that indeed CO flips have a much weaker interaction with the
phonon field compared to all other excitations studied, in agreement with the
two-TLS model, and similar to CN$^-$ flips in KBr:CN. However, because of the
untypical softness of the Ar:N$_2$:CO lattice, and the resulting smallness of
TLS-phonon interactions (both $\gamma_{\rm f}^{\rm CO}$ and $\gamma_{\rm
r}^{\rm CO}$ are smaller than $\gamma_{\rm f}^{{\rm CN}^-}$ and $\gamma_{\rm
r}^{{\rm CN}^-}$, by a factor of $\approx 3$ and by an order of magnitude,
respectively), we expect universality in Ar:N$_2$:CO to appear only below $T
\approx 0.1$ K. For Ar:N$_2$, where single impurity flips are absent, we find
a related absence of a distinct class of TLSs weakly interacting with the
strain. We then study the DOS of TLS bias energies for the different TLS
configurations, and find again no distinct class of TLSs typified by low bias energies.
Since Ar:N$_2$ does not fulfill the necessary conditions for universality as are
suggested by the two-TLS model\cite{SS09}, and in view of its low temperature thermal
conductivity as was obtained experimentally, we argue that Ar:N$_2$ constitutes a 
first example of a non-universal strongly disordered glass in three dimensions and 
in ambient pressure.

\section{Methods}
\label{Methods}

For most of our calculations, we have employed a model developed expressly for
this purpose. It is based on the iterative solution, using
previously described numerical methods\cite{kwon,burc}, of a mesh of non-linear
springs with Lennard-Jones $r^6-r^{12}$ potential\cite{lejo} to describe all
the interactions. In particular, we assumed the following Lennard-Jones parameters
$\epsilon_{Ar-Ar}=3.7936\cdot10^{-4} {\rm E_h}$, $\sigma_{Ar-Ar}=6.3302 r_B$,\cite{hoover}
$\epsilon_{Ar-N}=2.1263\cdot10^{-4} {\rm E_h}$, $\sigma_{Ar-N}= 6.3306 r_B$,\cite{nielaba}
$\epsilon_{N-N}= 0.3601 {\rm E_h}$, $\sigma_{N-N}= 2.0749 r_B$, \cite{raman},
$\epsilon_{N_2-N_2}=1.3916\cdot10^{-4} {\rm E_h}$, $\sigma_{N_2-N_2}= 6.3136
r_B$,\cite{johnson}, taking measures to deal with the numerical instabilities
that arise from the double N-N potential (intramolecular and intermolecular).

Starting from a pristine Ar lattice (a pure Ar network with crystallographic
positions\cite{nielaba}), we very gradually raise the concentration of $N_2$,
adding them one by one.  We do this up to a limit of 20\%Ar:80\%N$_2$. As
the Ar:$N_2$ ratio is decreased and to minimize artificial structural stress,
we adjust the lattice intersite distance to match the real density of Ar:N$_2$
mixtures.

In the N$_2$ enrichment process of the Ar lattice the goal is to minimize the
energy and at the same time to obtain a variety of structures so that the end
result is statistically representative. To achieve this, each new N$_2$
impurity added takes a random position and orientation, but every time that we
are about to add a new impurity, we choose up to 50 random positions and
orientations for it, starting from the same equilibrium lattice. For each of
these 50 possibilities, the structure of the lattice is relaxed, allowing the
positions of all atoms and the orientations of all molecules to adjust
slightly. Note that in the very first step, when the previous position was the
pristine Ar lattice, this generates 50 second-step structures, branching out
the procedure. Subsequent steps do not continue with the branching and instead
optimize the energy: for each one of the 50 structures with $n$ impurities we
obtain 50 with $n+1$, but keep only the most stable one.

We end up with 50 different low-energy configurations for each Ar:N$_2$ ratio.
For each of these, a Monte-Carlo procedure is used allowing for random
orientation tunneling of each N$_2$ molecule (25 sweeps) for further lowering
the energy. After each Monte-Carlo step the structure is relaxed. At the end of
the Monte-Carlo procedure, the system is ready to suffer the different kinds of
excitations as described below.

The efficiency of this model allows the use of rather large fragments, a
continuous exploration of the range of Ar:N$_2$ ratios and the obtention of
statistically significant number of low-energy configurations and all their
relevant excitations. We performed calculations both for 2D and 3D (hcp and
fcc) structures, but focused mainly on 2D, where reaching good low energy
states is easier because of the smaller number of stable N$_2$ orientations
($6$ in 2D vs. $12$ in 3D), and at the same time the constraints for finding
centrosymmetric TLSs are less stringent than in 3D, see below. More
details about the model can be found in a dedicated article.~\cite{vicente}

\begin{figure} [htb]
(a)\includegraphics[width=0.95\columnwidth]{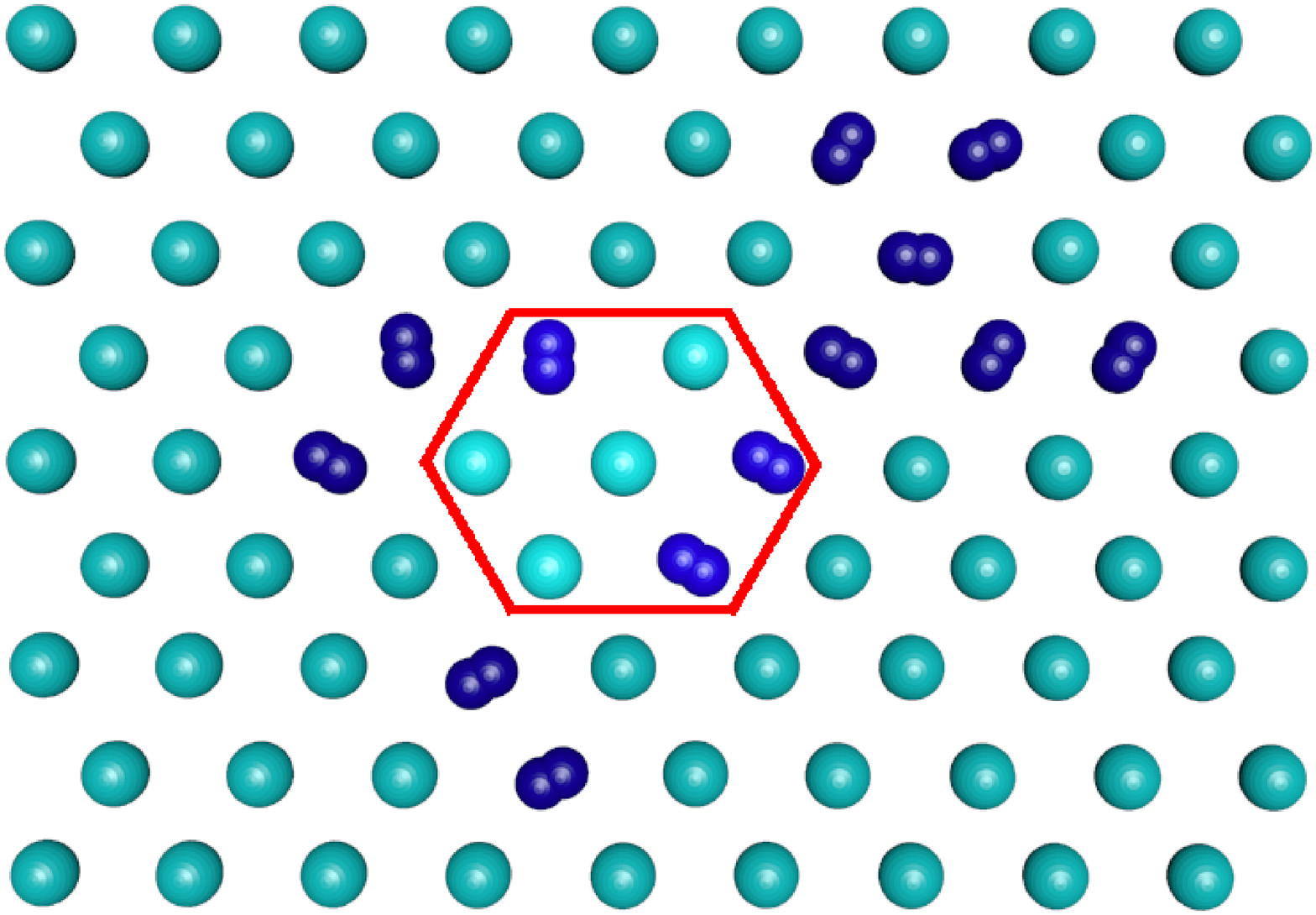}
\begin{tabular}{ccccc}
& & \\
%\hspace{-4.0cm}
(b)\includegraphics[width=0.27\columnwidth]{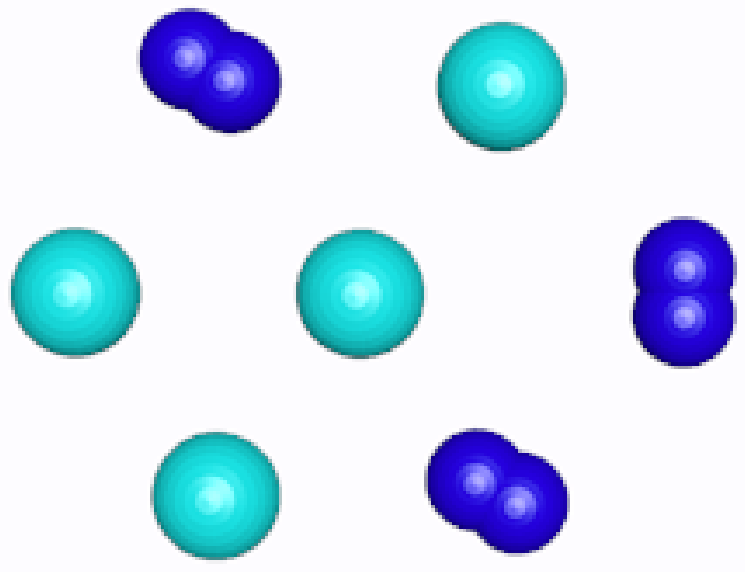} &
%\hspace{-4.0cm}
(c)\includegraphics[width=0.27\columnwidth]{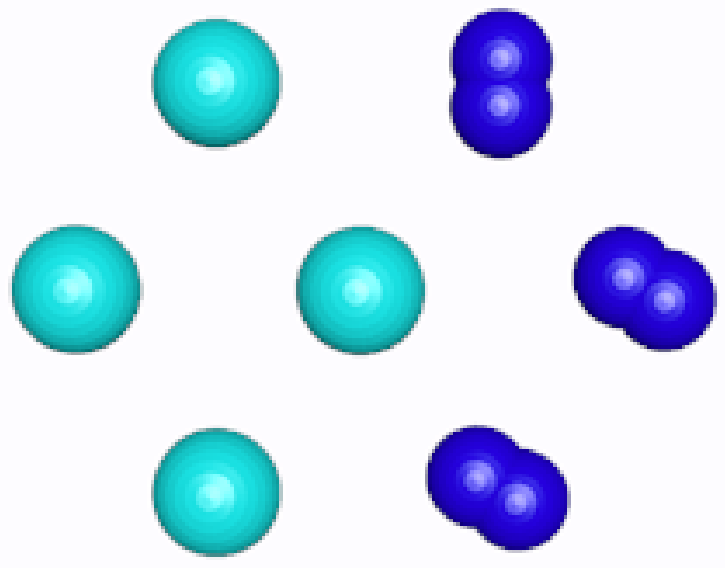} &
%\hspace{-4.0cm}
(d)\includegraphics[width=0.27\columnwidth]{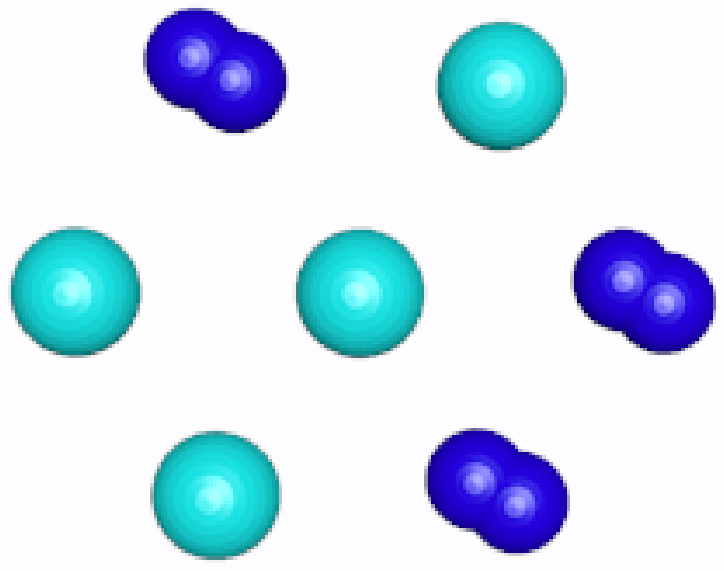} \\
\end{tabular}
\caption{(a) Initial configuration, highlighting the central hexagon where
excitations take place. (b)-(d) Different excited states starting from the same
initial state configuration. (b) flip-flop (c) Ar-tunneling (d) rotation.}
\label{excitations}
\end{figure}

To avoid border effects, we only study excitations within the seven central
positions of a 9x9 lattice (Fig. \ref{excitations} (a)). Within these seven
positions, we consider all possible N$_2$-N$_2$ (or Ar-N$_2$) pairs and the
three types of excitations depicted in Fig.~\ref{excitations} (b), (c), (d) and
detailed in the next section.

For the purposes of evaluating the density of states, we define the excitation
energy $\text{E$_{\rm bias}$}$ \eqref{eq:Edef} as the difference between the calculated
potential energies of the excited state $V_e$ and its corresponding ground
state $V_g$:
\begin{equation}
\label{eq:Edef}
\text{E$_{\rm bias}$}=\left(V_{e}-V_{g}\right) .
\end{equation}

To evaluate TLS-phonon coupling $\gamma$\eqref{eq:gamdef} we apply a 0.5\% mesh
contraction to obtain the difference between potential energies
\begin{equation}
\label{eq:gamdef}
\gamma=\left(V^{ph}_{g}-V_{g}\right)-\left(V^{ph}_{e}-V_{e}\right) \, .
\end{equation}
Here superscript $ph$ indicates the system after
the phonon and subscripts $g$ and $e$ indicate ground and excited state, respectively.
We have employed two different protocols, where in the first the system is relaxed 
after the mesh contraction and in the second it is not relaxed. Whereas single $\gamma$ values 
differ between these two protocols, differences between
the two protocols in the statistical distribution of $\gamma$ 
values as is plotted in Fig.~\ref{gammas} was found to be negligible.

For a few calculations on small 2D fragments, we employed the same DFT methods
as presented in~\cite{GS11} to extract the values of the TLS-phonon interaction
energy $\gamma$ of the different TLSs as explained above. We also use this
methodology to extract the values of the TLS-phonon interaction and estimate
$\gamma$ for CO head-tail flips (exactly analogous to the CN$^-$ flips
in~\cite{GS11}).

\section{Results}

As an initial exploration, we start from pristine Ar networks, either 2D or 3D,
and substitute N$_2$ for Ar at certain positions and in certain orientations.
Then we minimize the energy allowing only small displacements, and for each
starting configuration we arrive at a unique (local) ground state, in the sense
that we found no off-center displacement in any single N$_2$ molecule or in any
single Ar atom that could give rise to a centrosymmetric TLS.

On the other hand we did find pairs of low-energy configurations that relate to
each other through an inversion center, which we denote $\tau$-TLSs.  Those involve, in the
simplest case, the exchange of orientation of two non-parallel neighbouring
N$_2$, which we label as {\it flip-flop}.  Note that this is actually a
fragile, environmental-dependent phenomenon, where the low-energy part of a
complex spectrum of two or more non-centrosymmetric (denoted $S$-type) TLSs happens to
take the form of a $\tau-TLS$.  This is a fundamental difference with the
CN$^-$ case, where each molecule had an intrinsic, built-in, $\tau$-TLS in the
form of a CN$^-$ flip.
Because of this, the variable influence of an extra neighbouring N$_2$ results
in a wide spectrum for the interaction strength of flip-flop excitations
with the strain, ranging from $\gamma_{\rm f}$ to $\gamma_{\rm r}$.

A different, similarly fragile, $\tau$-TLS can be
defined if we consider a tunneling process where an Ar atom and an N$_2$
molecule exchange positions.  This we label as {\it Ar-tunnel}. Finally, any
change in orientation of a single given N$_2$ molecule always constitutes an
S-TLS. Except where otherwise stated, in our calculations we choose an
orientation change such that an N$_2$ adopts the orientation of a neighbouring,
non-parallel N$_2$, and label this as {\it rotation}. An example of each of
these three collective TLSs is shown in Figure~\ref{excitations}.

\begin{figure*}
\begin{tabular}{cc}
\includegraphics[width=0.450\textwidth]{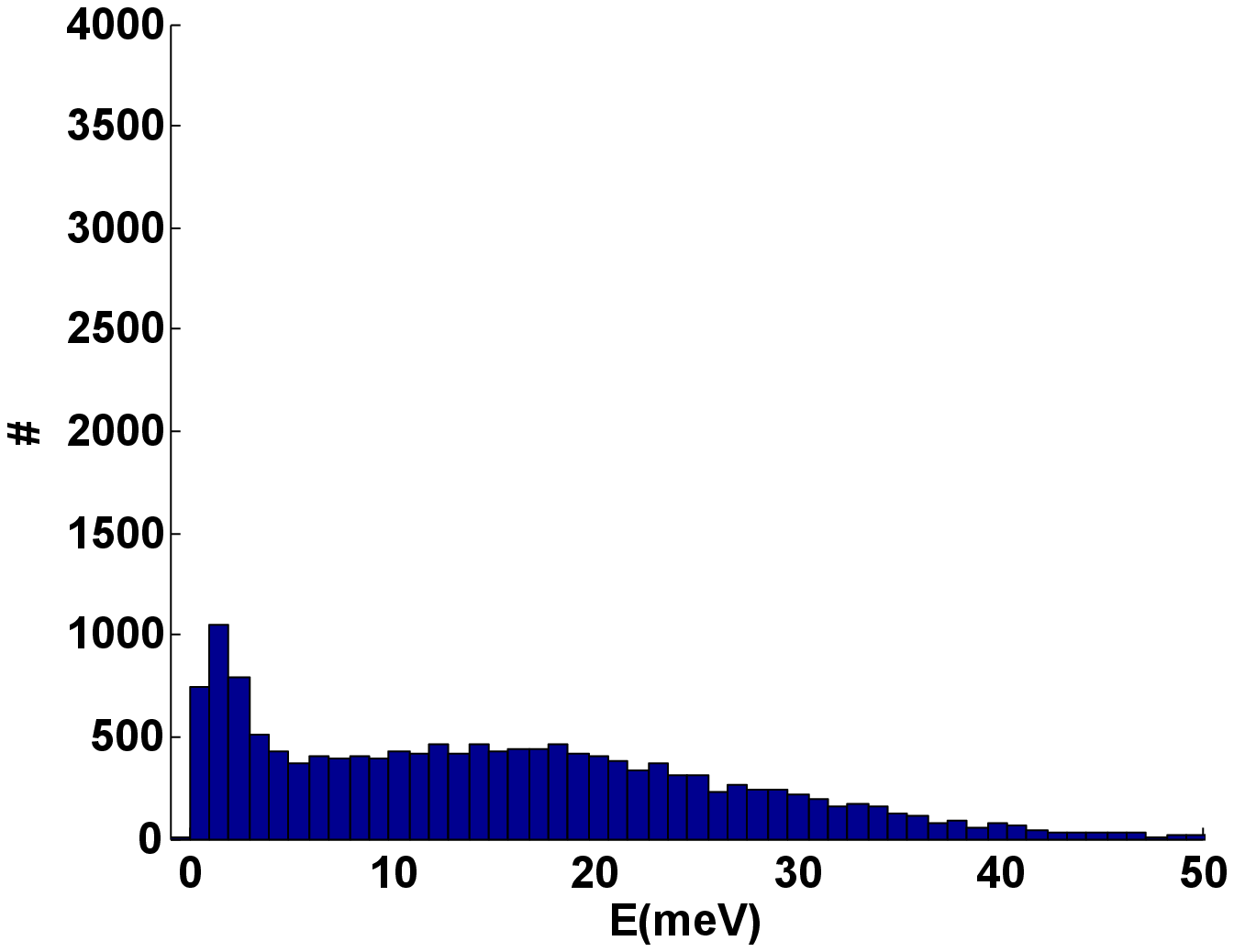}  &  \includegraphics[width=0.450\textwidth]{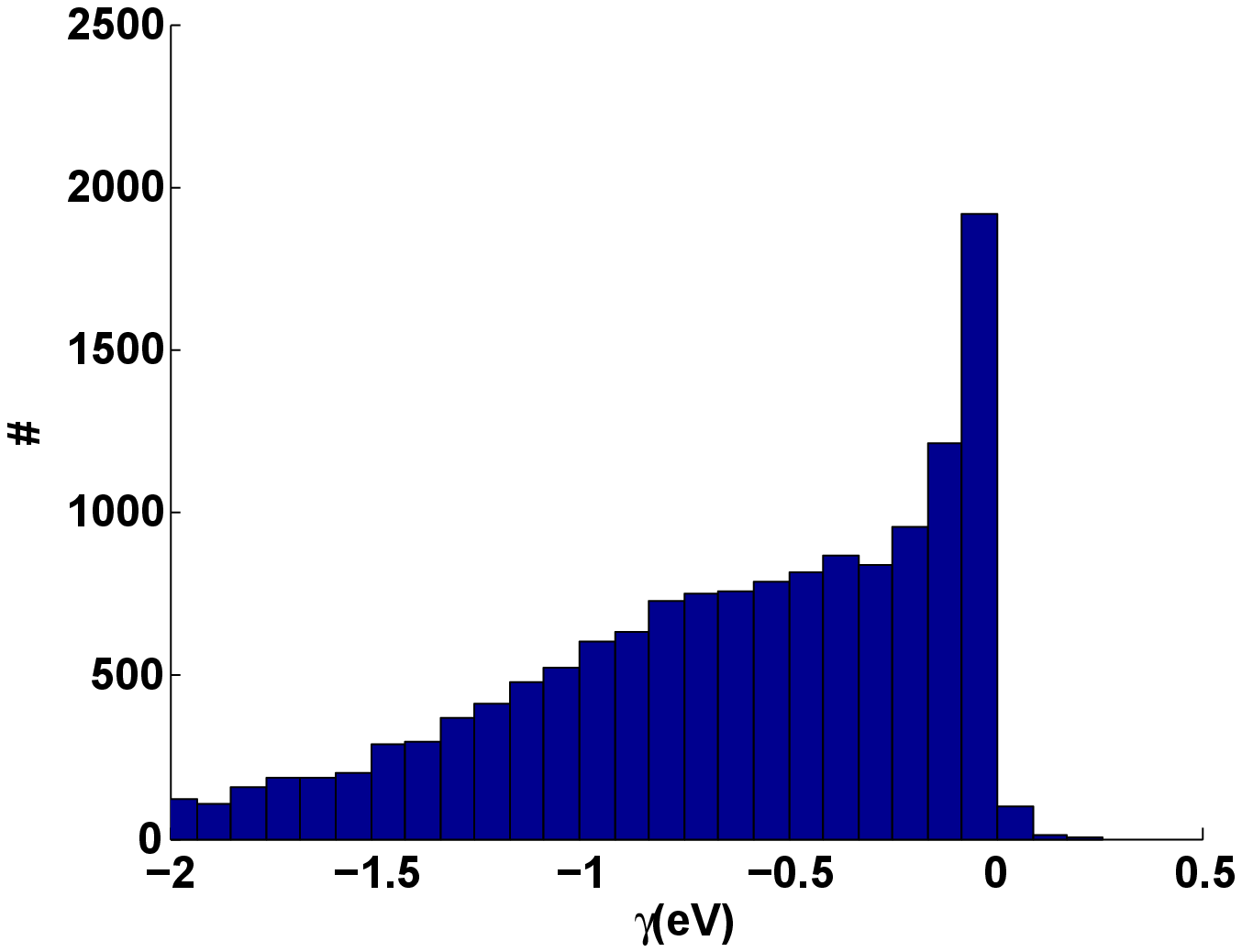}      \\
\includegraphics[width=0.450\textwidth]{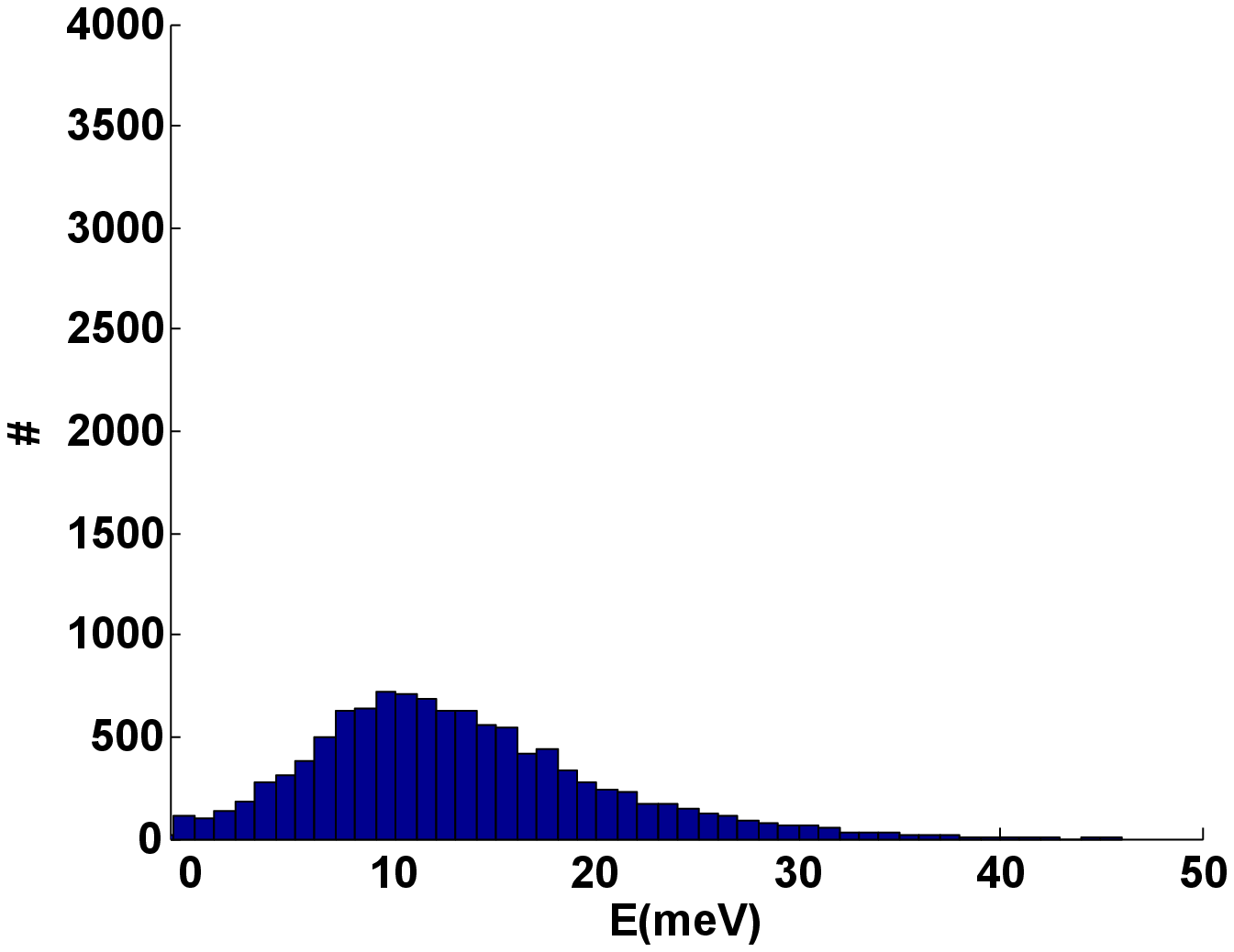}  &
\includegraphics[width=0.450\textwidth]{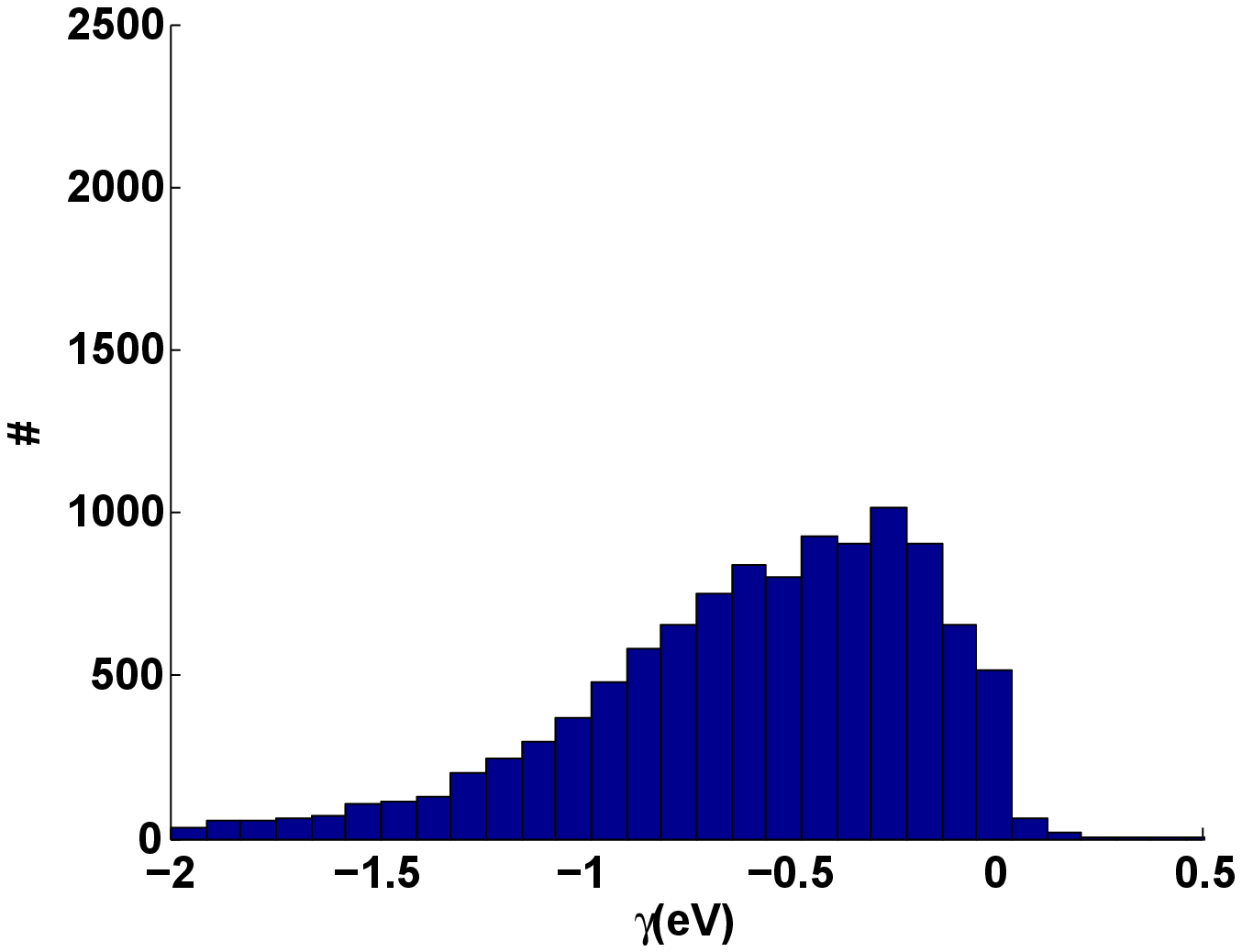}          \\
\includegraphics[width=0.450\textwidth]{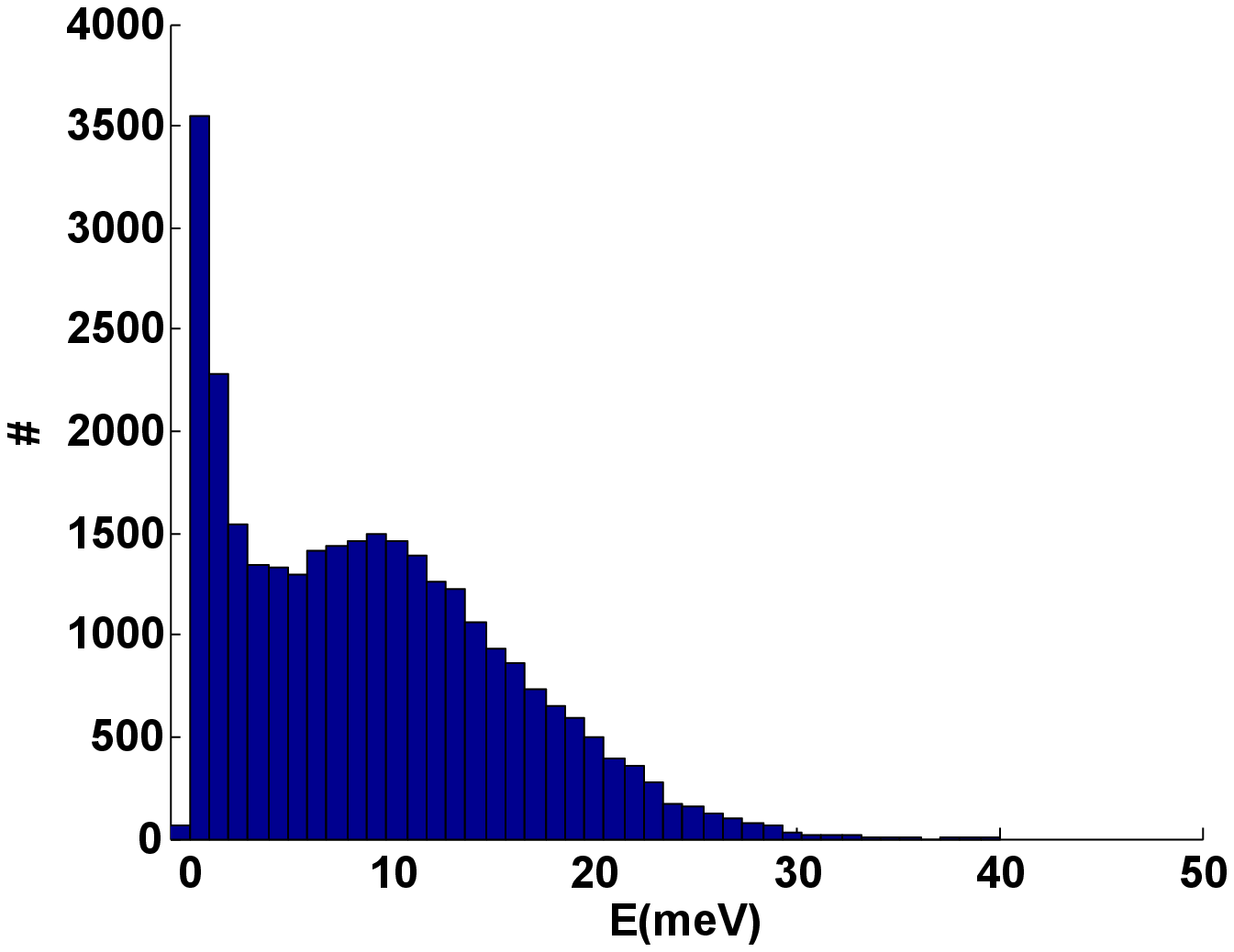}  &  \includegraphics[width=0.450\textwidth]{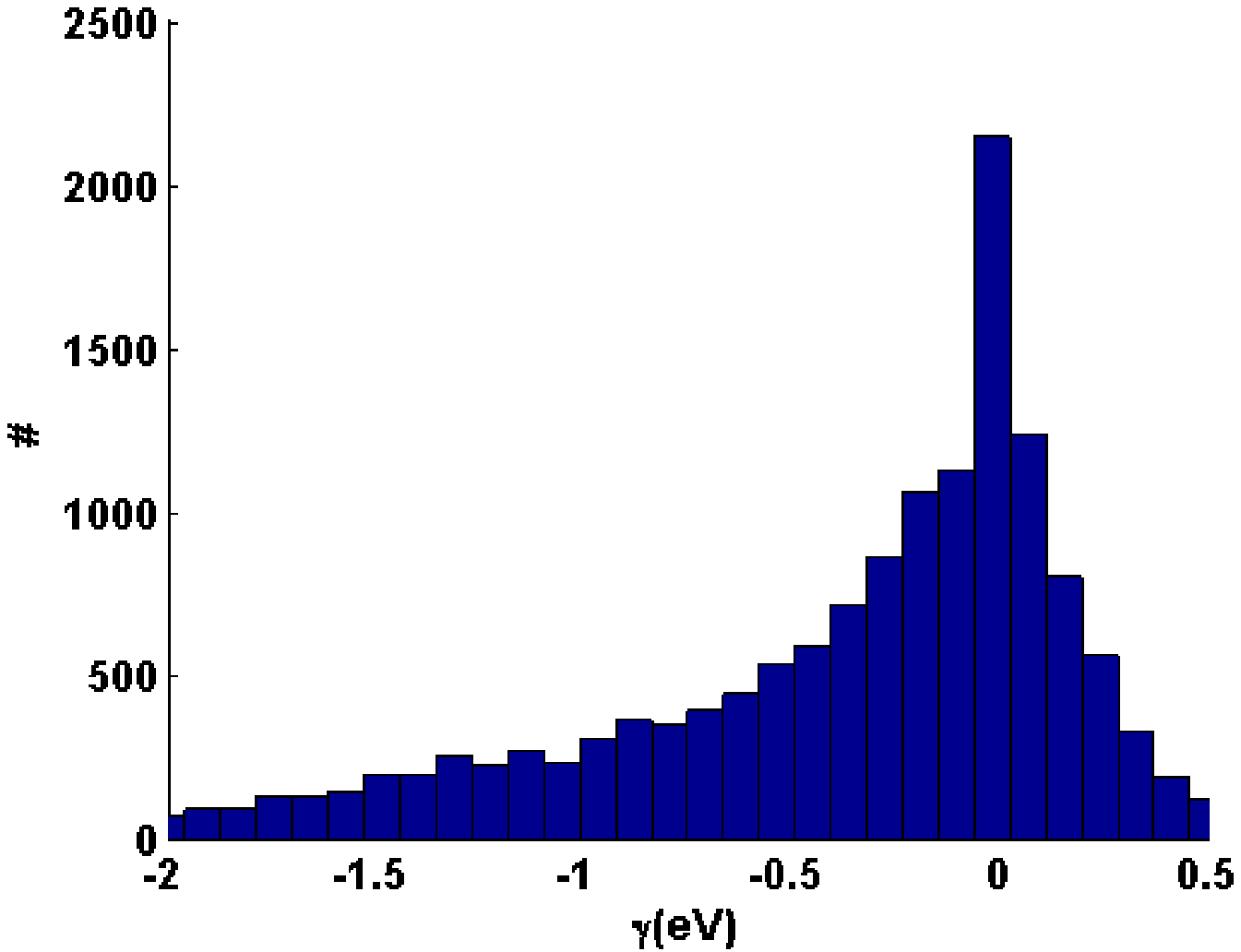} \\
\end{tabular}
\caption{Numerical results of 50 independent histories with Ar:N$_2$
ratios ranging from 0.8:0.2 to 0.2:0.8. Left: Density of states of bias energies 
[Eq.~(1)]. Right: frequency of encountered values for $\gamma$ [Eq.~(2)]. Exitation 
types from top to bottom: flipflop; Ar-tunnel; rotation [see Fig.~\ref{excitations}].}
\label{gammas}
\end{figure*}

To evaluate the character of these collective TLSs in realistic situations, we
proceed with a systematic exploration at increasing concentrations of N$_2$ as
detailed in section \ref{Methods} and in reference~\cite{vicente}.
As can be seen in Fig.~\ref{gammas}, we find for all TLSs distributions
of bias energies with typical values of $\gamma\approx 0.3-0.5$eV and
$E_{\rm bias} \approx10meV$, rather than markedly distinct behaviors for different TLS types
as would be expected for the $\tau$-$S$-TLSs model and was observed for KBr:CN
\cite{GS11}.  Of course, for every kind of TLS there are particular TLSs which
are almost uncoupled to phonons coming from a particular direction. Note that
in our model the phonon compression is not randomized, but is instead applied
in one of the special directions of the lattice. Therefore, it is expected that
there is a sub-class within each type of TLSs that presents low values of $\gamma$ to
phonons coming from a special direction.  The difference with the KBr:CN case
is the absence of two full (symmetry-defined) classes of TLSs $\tau$, $S$ where
the phonon coupling $\gamma$ is, for any direction, much lower in $\tau$ TLSs
compared with $S$ TLSs. The absence of such a class of TLSs is caused by
the dramatic perturbation of the
centrosymmetric TLSs by the nearest neighbours in the Ar:N$_2$ system. While
the interaction energies are the same in 2D and 3D, there are 12
nearest-neighbours in 3D compared with just 6 in 2D, meaning that the
centrosymmetric nature of collective TLSs will be, if at all, even more fragile
in 3D. We thus expect a similar wide distribution of the coupling
constants also for Ar:N$_2$ in three dimensions.

Notably, we also find that the obtained values for the TLS-strain 
interaction strengths 
are small compared with the $\gamma_{\rm r}$ determined for KBr:CN in a previous
work.\cite{GS11}. We have to emphasize that, up to a point, this was to be
expected. Indeed, N$_2$ and Ar (and CO, for that matter) are very different
from the vast majority of substances, in that they are formed by molecules that
are very weakly bounded to each other, compared with almost anything else. The
main interactions among these molecules are van der Waals rather than
electrostatic. Besides causing their extremely low boiling- and melting
temperatures, this also results in the low energy of their crystal defects. As
an example, the energy of a vacancy-type crystal defect in Ar is $< 0.1 {\rm
eV}$, while for iron --which is malleable-- this energy is $> 1{\rm eV}$.

%\begin{figure}[htb!]

Nevertheless, to further test this smallness of the $\gamma$, we construct
smaller lattices which we treat with the DFT methodology that
was originally used for the KBr:CN system \cite{GS11}. We employ the B3LYP functional and
the 6-311G basis set, plus a better basis set 6-311+G* to verify the
results. We confirm that the obtained $\gamma_{\rm r}$ are an order of magnitude
smaller than those obtained by the same methodology in KBr:CN (see
table~\ref{DFT-gammaS}).

\begin{table}[h]
\caption{$\gamma_{\rm r}$(rotation) in different configurations. Problem A: 3x3
pristine Ar lattice with a central N$_2$; A2 is the same lattice but using a
6-311+G* basis set.  Problem B: 5x5 pristine Ar lattice with a central N$_2$.
problem C: 5x5 pristine Ar lattice with three nearest neighbours N$_2$ in the
central line, oriented parallel to each other and perpendicular to the line
defined by their centers.}
\begin{tabular}{c|c|c|c|c}
                    &   A    &   A2   &   B    &   C   \\
\hline
$\gamma_{\rm r}$ (DFT,eV) &  0.60  &  0.56  &  0.320 &  0.22  \\
\end{tabular}
\label{DFT-gammaS}
\end{table}

We then used the
same methodology to estimate $\gamma$ for CO flips in Ar:N$_2$:CO samples.
we have found that $\gamma_{\rm f}$ is an order of magnitude smaller than $\gamma_{\rm r}$,
and that like
in the case of $\gamma_{\rm r}$, the values of $\gamma_{\rm f}$ are at least a
factor of three below the values found for flip excitations in KBr:CN
(Table~\ref{CO}). The tests were done on pristine 5x5 Ar fragments where some
of the Ar atoms in the central hexagon were substituted by N$_2$ molecules at
random orientations. In each configuration, we applied either horizontal or
vertical phonons.  For convenience and brevity, we label the N$_2$ impurities
according to their clock position and orientations.

\begin{table}[h]
\caption{$\gamma_{\rm f}$ for a CO impurity in the center of different 5x5
fragments. The low-state impurity is defined by its clock-orientation (e.g.
3=horizontal). All N$_2$ impurities are in the first hexagon and thus are
uniquely defined by their clock-position and -orientation, in that order.}
\begin{tabular}{c|c|c}
(CO)(N$_2$)(N$_2$)(N$_2$) & $\gamma_{\rm f,h}$ (meV) & $\gamma_{\rm f,v}$ (meV) \\
\hline
(3)(1,3);(5,3);(9,6)      &  31.8                   &  34.6                    \\
(3)(1,3);(3,4);(5,1)      &  33.7                   &  47.6                    \\
(5)(1,3);(3,3);(9,6)      &                         &   6.3                    \\
\end{tabular}
\label{CO}
\end{table}

\section{Discussion and conclusions}

KBr:CN is one of many disordered lattices showing low temperature phonon
attenuation properties which are equivalent in both functional form and
magnitude to those observed in all amorphous solids, and are thus dubbed
universal. For KBr:CN it was shown that universality is intimately related to
the existence of bi-modality in the typical values of the TLS-strain
coupling; inversion symmetric (flips, ``$\tau$") TLSs having a typically weak
interaction with the strain and inversion asymmetric (rotations, ``S") TLSs
having a typically strong interaction with the strain. The ratio between the
two interaction constants, $g \equiv \gamma_{\rm f}/\gamma_{\rm r}$, dictates
the ratio between the typical bias energies of $\tau$-TLSs and S-TLSs, and
consequently the universality and smallness of the tunneling strength. The
energy scale related to the temperature below which universality is observed is
given by the typical bias energy of the $\tau$-TLSs: $\gamma_{\rm f}
\gamma_{\rm r}/(\rho v^2 R_0^3)$, where $R_0$ is the typical distance between
impurities. This energy, being $g$ times smaller than the glass temperature, is
typically a few Kelvin.

As we found for the KBr:CN system\cite{GS11}, we find here also for the
Ar:N$_2$:CO system a bi-modality in the values of the interaction
constants of TLSs with the strain, where CO flips constitute a distinct
group of weakly interacting TLSs, {\it i.e.} $\gamma_{\rm f}$ has a typical
value much smaller than all other calculated TLS-phonon coupling constants,
including that of CO rotation. Thus, we expect the DOS of TLSs in Ar:N$_2$:CO 
to show a similar structure to that found in KBr:CN\cite{CBS14}, where the 
$\tau$-TLSs dominate the spectrum at low energies, have a roughly constant DOS, and thus dominate phonon attenuation. Furthermore, this structure of DOS coupled with the above mentioned bi-modality in the typical strengths of the TLSs interaction with the strain necessarily leads to qualitative and quantitative universality in phonon attenuation. We thus expect Ar:N$_2$:CO to show all the universal properties known in glasses.
However, since both $\gamma_{\rm f}$ and $\gamma_{\rm r}$ are untypically small for CO excitations in Ar:N$_2$:CO, we expect universality in this system to be present only at temperatures smaller than $\approx 0.1$ K.

The situation in Ar:N$_2$, in the absence of CO impurities, is very different,
since no single impurity flips exist.
Furthermore, we have shown here that no other excitation can play the role of such
flips in having a small typical interaction with the strain and small typical
bias energies in comparison to
all other excitations in this system; we find no off-center displacements of Ar
atoms or N$_2$ molecules; we then show explicitly that N$_2$ rotations and pair
excitations have strain interactions of typical value $\approx \gamma_{\rm r}$ 
and typical bias energies $\approx 10meV$. This is true also for pair excitations 
possessing local inversion symmetry, because the
proximity of each of the pair molecules (or Ar atom) to other N$_2$ molecules
in the solid yields large typical values for TLS-strain interactions and TLS
bias energies. Although we have not checked all types of excitations,
including those
involving larger number of impurities, we can not foresee a scenario in which
any such type of excitation would give rise to a systematically weak interaction with
the strain. Indeed, although excess low temperature specific
heat is found in Ar:N$_2$, resulting from an abundance of low energy excitations, its
thermal conductivity was found to have a very different temperature dependence than that
typical for glasses at low temperatures\cite{YNH89}. 
We therefore argue that Ar:N$_2$ is a non-universal glass, the first among
strongly disordered systems having tunneling states, whereas the apparent
similarity between the values of the specific heat in Ar:N$_2$ and Ar:N$_2$:CO\cite{NYHC87}
is limited to the relatively high temperature, in comparison to the energy scale
of the bias energies of CO flips, studied experimentally.

Our results here are tightly connected to the discussion regarding the
necessary and sufficient conditions to observe universality in disordered
lattices. In addition to strong strain disorder and tunneling
TLSs\cite{Wat95,TTP99}, the existence of a distinct class of TLSs weakly
interacting with the strain is required.
This is in line with the two-TLS model, showing that such a class of low energy
excitations is an outcome of the existence of inversion symmetric TLSs. Indeed,
we find that Ar:N$_2$, which does not show universal properties at low
temperatures, does not possess local inversion symmetric flip excitations, and
does not have a distinct class of weakly interacting TLSs. At the same time, we
refute the notion that CO flips in Ar:N$_2$:CO do not contribute to the low
energy properties of this system. In that we refute the central criticism to
the sufficiency of the above conditions,
%and in particular
including the presence of
inversion symmetric (flip) excitations, for the appearance of universality at
low temperatures.
Furthermore, our results suggest that the rather universal temperature
of $\approx 3$ K below which universality is observed is related also to the
fact that typical interactions between near neighbor impurities are rather similar in different
solids, Ar:N$_2$:CO being a marked exception.

Our conclusions here are based on the two-TLS model, and its proven
applicability to disordered lattices showing universality. Were these
conclusion to apply also to amorphous solids, it would suggest the generic
existence in amorphous solids of local excitations with markedly weak
interaction with the phonon field and small bias energies. It would
therefore be of much interest to check the applicability of the two-TLS model
to amorphous solids.

{\it Acknowledgments .---}
The present work has been funded by the EU (Project ELFOS and ERC Advanced
Grant SPINMOL), the Spanish MINECO (grant MAT2011-22785, the CONSOLIDER
project on Molecular Nanoscience), the Generalitat Valenciana (Prometeo
and ISIC Programmes of excellence) and the Israel Science Foundation (Grant No.
982/10). A.G.A. acknowledges funding by the MINECO (Ram\'on y Cajal Program).

\end{document}